\documentstyle[preprint,prc,aps]{revtex}

\begin{document}
\draft

\title{$\phi$-meson production in proton-proton collisions}

\author{K. Nakayama$^{a,b}$, J. W. Durso$^{a,c}$, J. Haidenbauer$^a$,
C. Hanhart$^{a,d,e}$, and J. Speth$^a$}

\address{$^a$Institut f\"{u}r Kernphysik, Forschungszentrum J\"{u}lich
GmbH, D--52425 J\"{u}lich, Germany \\
$^b$Department of Physics and Astronomy, University of Georgia, Athens, GA 
30602, USA \\
$^c$Physics Department, Mount Holyoke College, South Hadley, MA 01075, USA \\
$^d$Institut f\"{u}r Theoretische Kernphysik, Universit\"{a}t Bonn, D-53115 
Bonn, Germany \\
$^e$Present address: Department of Physics and INT, University of Washington, \\
Seattle, WA 98195, USA}

\maketitle

\begin{abstract}
The production of $\phi$-mesons in proton-proton collisions is investigated 
within a relativistic meson-exchange model of hadronic interactions. The 
experimental prerequisites for extracting the $NN\phi$ coupling strength from
this reaction are discussed. In the absence of a sufficient set of data, which
would enable an accurate determination of the $NN\phi$ coupling strength, we 
perform a combined analysis, based on some reasonable assumptions, of the 
existing data for both $\omega$- and $\phi$-meson production. We find that the
recent data from the DISTO collaboration on the angular distribution of the 
$\phi$ meson indicate that the $NN\phi$ coupling constant is small. The 
analysis yields values for $g_{NN\phi}$ that are compatible with the OZI rule.
\end{abstract}


\newpage

%----------------------
 \section{Introduction}
%----------------------

The investigation of $\phi$-meson production in both hadronic
\cite{Ast,CrB,Obe,Ber95,Wur96,Disto,Ell95,Gut97,Loche,Buzat,Klemp,Mul94,Rekal,Tito1,Kondr}
and electromagnetic \cite{Willi,Tito2} processes at low and
intermediate energies has attracted much attention in recent years. A
major motivation for those studies is the hope that one can obtain
information about the amount of hidden strangeness in the
nucleon. Indications for a possibly significant $\bar ss$ component in
the nucleon have been found, among others, in the analysis of the
so-called $\pi N$ $\Sigma$ term \cite{Don86} and from the EMC
measurement of deep inelastic polarized $\mu p$ scattering
(``nucleon-spin crisis'') \cite{EMC88}.

In the context of $\phi$-meson production processes one expects 
\cite{Ell89,Hen92} that a large amount of hidden strangeness in the nucleon 
would manifest itself in reaction cross sections that significantly exceed the
values estimated from the OZI rule \cite{OZI}. This phenomenological rule 
states that reactions involving disconnected quark lines are forbidden. In the
naive quark model the nucleon has no $\bar ss$ content, whereas the 
$\phi$-meson is an ideally mixed pure $\bar ss$ state. Thus, in this case, the
OZI rule implies vanishing nucleon-nucleon-$\phi$ ($NN\phi$) coupling and, 
accordingly, a negligibly small production of $\phi$-mesons from nucleons (or
anti-nucleons) by electromagnetic or (non-strange) hadronic probes.  In 
practice there is a slight deviation from ideal mixing of the vector mesons, 
which means that the $\phi$-meson has a small $\bar uu + \bar dd$ component. 
Thus, even if the OZI rule is strictly enforced, there is a non-zero coupling
of the $\phi$ to the nucleon, although the coupling is very small. Its value 
can be used to calculate lower limits for corresponding cross sections. For 
example, under similar kinematic conditions to cancel out phase space effects, 
one expects cross section ratios of reactions involving the production of a 
$\phi$- and an $\omega$-meson, respectively, to be \cite{Lip76}
\begin{equation}
R = {\sigma (A+B\rightarrow \phi X)
  \over \sigma (A+B\rightarrow \omega X)}  \approx \tan ^2 (\alpha_V)\ ,
\label{Ratio}
\end{equation}
where $A$, $B$ and $X$ are systems that do not contain strange quarks. Here, 
$\alpha_V \equiv \theta_V - \theta_{V(ideal)}$ is the deviation from the ideal
$\omega - \phi$ mixing angle. With the value $\alpha_V \cong 3.7^o$ \cite{PDG}
one gets the rather small ratio of $R = 4.2 \times 10^{-3}$. 
 
Recent experiments on antiproton-proton ($\bar pp$) annihilation at rest at 
the LEAR facility at CERN revealed, however, considerably larger branching
ratios for various $\phi$ production channels \cite{Ast,CrB,Obe}. 
Many of the measured ratios $\sigma(\bar pp \rightarrow \phi X) / 
\sigma(\bar pp \rightarrow \omega X)$ are about $100 \times 10^{-3}$ 
or even larger (cf. the compilation of data given in Ref.~\cite{Ell95}), which 
means that they exceed the estimate from the OZI rule by more than one order
of magnitude. Significant deviations from the OZI rule were also found
in the reactions $\bar p p\rightarrow \phi\phi$ \cite{Ber95} and 
$pd \rightarrow ^3$He$\phi$ \cite{Wur96}. 

These observed large $\phi$-production cross sections were interpreted by some 
authors as a clear signal for an intrinsic $\bar ss$ component in the nucleon 
\cite{Ell95,Gut97}. However, this explanation is still controversial. There is 
an alternative approach aimed at understanding these large cross sections solely 
by the strong rescattering effects in the final state \cite{Loche,Buzat,Klemp,Mul94}.
In this case $\phi$ production occurs via two-step processes with intermediate 
states such as $\bar KK$, $\bar KK^*$, $\bar \Lambda \Lambda$, etc., where each 
step is allowed by the OZI rule. Corresponding quantitative calculations have, 
indeed, demonstrated that the resulting production cross sections are sufficiently 
large to agree with the experiments without a violation of the OZI rule 
\cite{Loche,Buzat,Klemp,Mul94}.  

In this connection, the production of $\phi$-mesons in proton-proton ($pp$) 
collisions is of special interest. In principle, the production cross section can 
be used for a direct determination of the $NN\phi$ coupling strength. Any
appreciable $NN\phi$ coupling in excess of the value given by the OZI rule could 
be seen as evidence for hidden strangeness in the nucleon. Of course, there is also 
an alternative picture: one in which the coupling of the $\phi$-meson to the nucleon 
does not occur via possible $\bar ss$ components in the nucleon, but via intermediate 
states with strangeness \cite{Geiger,Meissner}. Specifically, this means that the 
$\phi$-meson couples to the nucleon via virtual $\Lambda K$, $\Sigma K$, etc. states. 
Corresponding model calculations have shown, however, that such processes give rise 
to (effective) $NN\phi$ coupling constants comparable to the OZI values and therefore 
should not play a role in drawing conclusions concerning hidden strangeness in the 
nucleon. Accordingly, one expects that cross section ratios 
$\sigma(pp\rightarrow pp\phi)/ \sigma(pp\rightarrow pp\omega)$ should provide a clear 
sign for a possible OZI violation. Indeed, data presented recently by the DISTO 
collaboration \cite{Disto} indicate that this ratio, after correcting for the phase 
space effects, is about eight times larger than the OZI estimate (\ref{Ratio}). 

The present work focuses on the $pp \rightarrow pp\phi$ process. 
We discuss the experimental prerequisites that would enable 
one to disentangle competing reaction mechanisms for $\phi$-production
in nucleon-nucleon ($NN$) collisions and to extract the value of the $NN\phi$ 
coupling strength. In this context we shall show that the angular 
distribution of the produced $\phi$-mesons plays a crucial role.
In particular, we will demonstrate that the almost isotropic angular 
distribution seen in the data from the DISTO collaboration \cite{Disto} 
indicates that the $NN\phi$ coupling is very small. In the absence of a set of 
data sufficient to permit an accurate determination of the $NN\phi$ coupling 
strength, we impose certain reasonable assumptions and carry out a 
combined analysis of the recent $\phi$-meson production data of the 
DISTO collaboration \cite{Disto} and the data of $\omega$-meson production from 
SATURNE \cite{Saclay}. This analysis yields a range of values for $g_{NN\phi}$ 
that is compatible with the OZI rule. 

We describe the $pp \rightarrow pp\phi$ reaction within a 
relativistic meson-exchange model. 
(See Ref.~\cite{Nak1} for the details of the formalism.) 
The transition amplitude is calculated in 
Distorted Wave Born Approximation, where the $NN$ final state interaction 
is taken into account explicitly. The final state interaction is known to be 
very important at near-threshold energies.   
For the $NN$ interaction we employ the model Bonn B as defined
in Table A.1 of Ref.~\cite{MHE87}. This model reproduces the $NN$ phase shifts
up to the pion-production threshold as well as the deuteron 
properties \cite{MHE87}.  As in our previous work \cite{Nak1}, we do not 
consider the initial state interaction explicitly.  
At the corresponding high incident energies the $NN$ interaction 
is a slowly varying function of energy.  Its main effect is to lead to an
overall reduction of the $pp\rightarrow pp\phi$ cross section \cite{Hanhart}, 
as has been shown explicitly for the case of the reaction
$pp \rightarrow pp\eta$ by Batini\'c et al. \cite{BSL}.
In the present model this effect of the initial state interaction is accounted 
for by an appropriate adjustment of the (phenomenological) form factors at 
the hadronic vertices.

In the next section we discuss possible basic production currents that can
contribute to the reaction $pp\rightarrow pp\phi$. Using SU(3) flavor
symmetry and imposing the OZI rule, we calculate meson-meson-meson and
nucleon-nucleon-meson coupling constants that are relevant for the 
$\phi$-production currents
and then estimate the corresponding contributions to the total cross
section. These reveal that the $\phi\rho\pi$ meson-exchange current is
the dominant one. 
In the same section we also give a short outline of our formalism and
introduce the free parameters of our model.  
In Sect.\ III we describe in detail the procedure for the combined 
analysis of $\phi$- and $\omega$-production data.  
We conclude with a summary of our results in the last section.

%------------------------------
 \section{Production currents}
%------------------------------

\begin{figure}[h]
\vspace{6cm}
\includegraphics{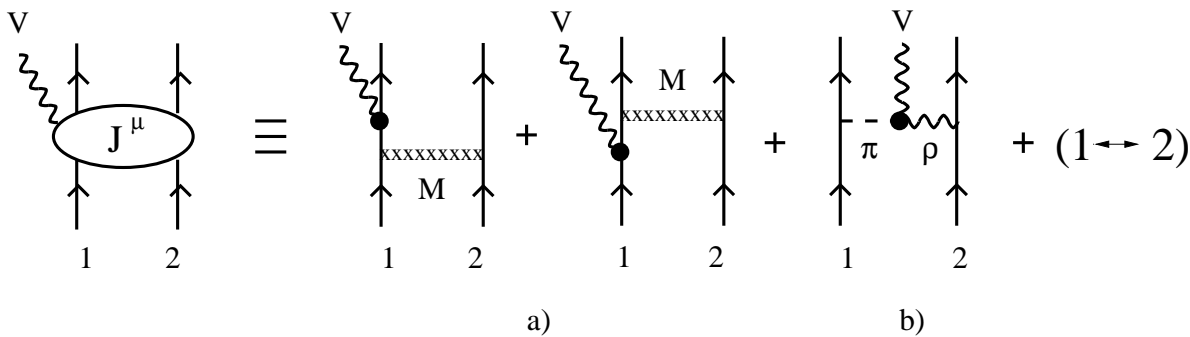}
\caption{$\phi$ and $\omega$-meson production currents, $J^\mu$, included in 
the present study: (a) nucleonic current, (b) meson exchange current. 
$v = \omega, \phi$ and $M = \pi, \eta, \rho, \omega, \sigma, a_o$.}
\label{fig1}
\end{figure}

In the case of the reaction $pp\rightarrow pp\omega$, the dominant production 
mechanisms
---namely, the nucleonic and $\omega\rho\pi$ mesonic currents, as depicted in 
Fig.~\ref{fig1}---can be identified without involved considerations. 
This is because of the 
relatively large $NN\omega$ and $\omega\rho\pi$ coupling strengths. The
$\omega\rho\pi$ meson-exchange current depends also on 
the $NN\rho$ and $NN\pi$ couplings, which are likewise large. Therefore
this production mechanism is by far the dominant one among the 
possible exchange currents \cite{Nak1}. 
For $\phi$-meson production, however, the situation is much less 
clear {\it a priori} and requires careful consideration. 
In order to determine the importance of various
possible meson-exchange currents, we first estimate systematically the coupling 
strengths of the hadronic vertices that enter into these
$\phi$-production exchange currents. This is done using effective 
Lagrangians, assuming SU(3) flavor symmetry, and imposing the OZI rule. 
We note that in this scheme the SU(3) symmetry breaking is introduced through 
the use of the physical masses of the hadrons in calculating the observables. 

In the present work we are interested in three-meson vertices involving at least 
one $\phi$-meson. We, then, have $VVP$, $VPP$ and $VVV$ vertices ($V =$ vector 
meson, $P =$ pseudo-scalar meson). The SU(3) effective Lagrangian has the form  
\begin{eqnarray}
\nonumber
{\cal L}_{MMM} & = & g_{888}\left[-(1-\beta)Tr([M_8,M_8]M_8) + 
                                \beta Tr(\{M_8,M_8\}M_8) \right] \\ \nonumber
& + & g_{88s}\beta Tr(\{M_8,M_8\}M_s) +    g_{8s8}\beta Tr(\{M_8,M_s\}M_8) \\
& + & g_{sss}\beta Tr(\{M_s,M_s\}M_s) \ ,
\label{SU3LM}
\end{eqnarray}
where $M_8$ ($M_s$) stands for the SU(3) meson-octet (-singlet) matrix,
and $\beta$ is the parameter specifying the admixture of the $D$-type 
($\beta =1$) and $F$-type ($\beta = 0$) couplings. The OZI condition relates 
the basic coupling constants of the SU(3) octet and singlet members in the 
Lagrangian, $g_{sss}=g_{88s}=g_{8s8}=\sqrt{{2\over 3}} \ g_{888}$, 
so that there is only one free coupling constant. This one is then fixed 
by a fit to some appropriate observables such as decay rates. 
The $VPP$ and $VVV$ vertices involve only 
$F$-type coupling if we require charge conjugation invariance. This leads to 
the vanishing of all $VPP$ and $VVV$ couplings involving either the $\omega$- 
or $\phi$-meson relevant to the present work. 
The $D$-type coupling leads to G-parity violating 
vertices and will be ignored. In contrast, the $VVP$ vertices involve 
only $D$-type coupling. In Ref.~\cite{Durso} essentially the same scheme 
as proposed here has been used to describe radiative meson decays. 
Therefore, we use the model parameters determined in that paper. 
The effective Lagrangian used in 
Ref.~\cite{Durso} can easily be cast into the form given by Eq.(\ref{SU3LM}). 
The resulting coupling constants $g_{VVP}$ at those
vertices relevant for the meson-exchange currents of interest
are tabulated in Table~\ref{tab1}. 

In order to estimate the magnitude of the various meson-exchange currents 
one needs not only the $VVP$ couplings, but the corresponding $NNV$ and 
$NNP$ couplings as well. Thus, in addition to the $NN\pi$ and $NN\rho$ couplings, the 
empirically poorly known $NN\omega$, $NN\phi$, $NN\eta$ and $NN\eta '$ couplings
are needed. We estimate these coupling strengths using the same scheme used for 
determining the $VVP$ couplings, i.e., the SU(3) effective Lagrangian
\begin{eqnarray}
\nonumber
{\cal L}_{BBM} & = & g_8\left[-(1-\beta)Tr([\bar B,B]M_8) 
                                + \beta Tr(\{\bar B,B\}M_8) \right] \\
               & + & g_s \beta  Tr(\{\bar B,B\}M_s) \ ,
\label{SU3LB}
\end{eqnarray}
plus the OZI rule. This latter condition yields $g_s = (3-4\beta)g_8/\sqrt{6}$. 
In the above Lagrangian $B$ stands for the SU(3) baryon-octet matrix. The 
$D$-type coupling in the SU(3) Lagrangian is not allowed for the $NNV$ 
couplings if one requires spin-independence for $BB\omega$ 
and $BB\phi$ couplings within the identically flavored baryons, 
$B=\Sigma, \Lambda$. Using the value of the $\omega - \phi$ mixing angle 
from Ref.~\cite{PDG} and $g_8 (= g_{NN\rho})$ from 
Ref.~\cite{MHE87}, we find $g_{NN\omega} \cong 9$ and 
$g_{NN\phi} \cong -0.6$ for the $NN\omega$ and $NN\phi$ vector-coupling 
constants, respectively. The $NNv$ tensor-coupling constants are simply 
chosen to be $f_{NNv} = \pm 0.5 \ g_{NNv}$, where $v \equiv \omega, \ \phi$ 
(as distinct from $V$, which stands for vector mesons in general).    
This choice of $f_{NNv}$ is consistent with values from an SU(3) 
estimate \cite{Dover} 
as well from other sources \cite{Meissner,MHE87}. 
For the $NN\eta$ coupling constant we take the value used 
in $NN$ scattering analysis \cite{MHE87}, $g_{NN\eta}=6.14$. This value, 
together with the $\eta - \eta'$ mixing angle of $\theta_P \cong -9.7^o$, 
as suggested by the quadratic mass formula, and the $NN\pi$ coupling 
constant, $g_{NN\pi}= 13.45$, leads to the ratio 
$D/F \equiv \beta / (1 - \beta) \cong 1.43$ -- which 
is close to the value of $D/F = 1.58 \pm 0.07$ extracted from a 
systematic analysis of semileptonic hyperon decays \cite{Bourquin}. 
The $NN\eta'$ coupling constant is related to the $NN\eta$ coupling 
constant by $g_{NN\eta'} = - g_{NN\eta} \tan(\alpha_P)$, where 
$\alpha_P \equiv \theta_P - \theta_{P(ideal)} \cong -45^o$ 
denotes the deviation from the pseudoscalar ideal mixing angle. 
We find $g_{NN\eta'} \cong 6.1$.  
 
Once all the relevant coupling constants have been determined, we can estimate the 
relative importance of various $VVP$-exchange currents to the $\phi$-meson
production. In corresponding test calculations it turned out that 
the $\phi\rho\pi$-exchange current is by far the  
dominant mesonic current. The {\it combined} contribution of all other 
exchange currents to the total cross section is about two orders of magnitude 
smaller. 

We have also examined contributions from meson-exchange currents
involving other heavy mesons, in particular, the $\phi\phi f_1$- and 
$\phi\omega f_1$-exchange currents, whose coupling constants (upper limits) can be
estimated from the observed decay of $f_1 \rightarrow \phi + \gamma$ \cite{PDG}.
We find a negligible contribution to the $\phi$-meson production. 
Furthermore the $\phi\phi\sigma$- and $\phi\omega\sigma$-exchange 
currents also turned out to be negligible.
Finally we note that, as in the case of $\omega$ production, there are no 
experimental indications of any of the known isospin-1/2 $N^*$ resonances 
decaying into $N\phi$. 

With the above considerations, the $v$-meson ($v=\omega, \phi$) production 
current 
$J^\mu$ is given by the sum of the nucleonic and $v\rho\pi$ meson-exchange currents, 
$J^\mu = J^\mu_{nuc}  + J^\mu_{mec}$, as illustrated diagrammatically in 
Fig.~\ref{fig1}. Explicitly, the nucleonic current is defined as
\begin{equation}
J^\mu_{nuc} = \sum_{j=1,2}\left ( \Gamma^\mu_j iS_j U + U iS_j \Gamma^\mu_j 
\right ) \ ,
\label{nuc_cur}
\end{equation}
with $\Gamma^\mu_j$ denoting the $NNv$ vertex and $S_j$ the nucleon (Feynman) 
propagator for nucleon $j$. The summation runs over the two interacting 
nucleons, 1 and 2. $U$ stands for the meson-exchange $NN$ potential. It is, in 
principle, identical to the driving potential used in the construction of 
the $NN$ interaction \cite{MHE87}, except that here meson retardation effects
are retained following the Feynman prescription. Eq.(\ref{nuc_cur}) 
is illustrated in Fig.~\ref{fig1}a.

The structure of the $NNv$ vertex, $\Gamma^\mu_j$, required in Eq.(\ref{nuc_cur}) 
for the production is obtained from the Lagrangian density
\begin{equation}
{\cal L}(x) =  - \bar\Psi(x) \left( g_{NNv}
       [\gamma_\mu -{\kappa_v\over 2m_N}\sigma_{\mu\nu}\partial^\nu ] V^\mu(x)
             \right) \Psi(x) \ ,
\label{NNv}
\end{equation}
where $\Psi(x)$ and $V^\mu(x)$ stand for the nucleon and vector-meson field,
respectively. $g_{NNv}$ denotes the vector coupling constant and $\kappa_v 
\equiv f_{NNv}/g_{NNv}$, with $f_{NNv}$ the tensor coupling constant. 
$m_N$ denotes the nucleon mass.

As in most meson-exchange models of interactions, each hadronic 
vertex is furnished with a form factor in order to account for, among other 
things, the composite nature of the hadrons involved. In this spirit
the $NNv$ vertex obtained from the above Lagrangian 
is multiplied by a form factor. 
The theoretical understanding of this form factor is beyond the 
scope of the present paper; we assume it to be of the form
\begin{equation}
F_{vNN}(l^2) = \frac{\Lambda_{Nv}^4} {\Lambda_{Nv}^4 + (l^2-m_N^2)^2}  \ ,
\label{formfactorN}
\end{equation}
where $l^2$ denotes the four-momentum squared of either the incoming or
outgoing off-shell nucleon. It is normalized to unity when the nucleon is
on its mass shell, i.e., when $l^2 = m_N^2$.

The $v\rho\pi$ vertex required for constructing the meson-exchange current,
$J^\mu_{mec}$ (Fig.~\ref{fig1}b), is derived from the Lagrangian density
\begin{equation}
{\cal L}_{v\rho\pi}(x) = \frac{g_{v\rho\pi}} {\sqrt{m_v m_\rho}}
\varepsilon_{\alpha\beta\nu\mu} \partial^\alpha \vec \rho^\beta(x) \cdot
\partial^\nu \vec \pi(x) V^\mu(x) \ ,
\label{pirhoomega}
\end{equation}
where $\varepsilon_{\alpha\beta\nu\mu}$ denotes the Levi-Civita antisymmetric
tensor with $\varepsilon_{0123}=-1$. The $v\rho\pi$ vertex obtained from the 
above Lagrangian is multiplied by a form factor which is taken to be of the
form
\begin{equation}
F_{v\rho\pi}(q_\rho^2, q_\pi^2) = 
\left ( \frac{\Lambda_{Mv}^2 - x m_\rho^2} {\Lambda_{Mv}^2 - q_\rho^2} \right ) 
\left ( \frac{\Lambda_{Mv}^2 - m_\pi^2} {\Lambda_{Mv}^2 - q_\pi^2} \right ) \ . 
\label{formfactorM}
\end{equation}
It is normalized to unity at $q_\pi^2 = m_\pi^2$ and $q_\rho^2 = x m_\rho^2$.
The parameter $x$ (= 0 or 1) is introduced in order to allow for different
normalization points as explained later. The form factor given above differs 
from the one used in \cite{Nak1} in that it uses a common cutoff 
parameter $\Lambda_{Mv}$ for both $\pi$ and $\rho$-meson instead of separate
cutoff masses. 

The meson-exchange current is then given by
\begin{equation}
J^\mu_{mec} = [\Gamma^\alpha_{NN\rho}(q_\rho)]_1 iD_{\alpha\beta}(q_\rho)
              \Gamma^{\beta\mu}_{v\rho\pi}(q_\rho, q_\pi, k_v)
              i\Delta(q_\pi) [\Gamma_{NN\pi}(q_\pi)]_2   +  
(1\leftrightarrow 2) \ ,
\label{mec_cur}
\end{equation}
where $D_{\alpha\beta}(q_\rho)$ and $\Delta(q_\pi)$ stand for the $\rho$- and 
$\pi$-meson (Feynman) propagators, respectively. The vertices $\Gamma$ 
involved are self-explanatory. Both the $NN\rho$ and $NN\pi$ vertices,
$\Gamma^\alpha_{NN\rho}$ and $\Gamma_{NN\pi}$, are taken consistently with
the $NN$ potential used to generate the $NN$ final state interaction.

Our model for vector-meson production described above contains five
parameters: two for the mesonic current (the coupling constant 
$g_{v\rho\pi}$ and the cutoff parameter $\Lambda_{Mv}$) and three for the 
nucleonic current (the coupling constants $g_{NNv}$ and $f_{NNv}$, and the 
cutoff parameter $\Lambda_{Nv}$). 
In Ref.~\cite{Nak1} we pointed out that the angular distribution of the 
emitted $\omega$-mesons is a sensitive quantity for determining the 
{\it absolute amount} of nucleonic as well as mesonic current 
contributions, in addition to the relative sign between the two contributions.
This applies also to the case of $\phi$-meson production since, as we have 
argued above, the dominant production mechanisms are exactly the same for 
both processes. Specifically, this means that the knowledge of the
angular distribution allows one to fix the cutoff parameter in the mesonic current 
since the coupling constant $g_{v\rho\pi}$ can be extracted from the 
relevant measured partial decay widths \cite{PDG}.
    
The nucleonic current, however, involves three free parameters. 
In this case knowledge of the angular distribution 
allows one to determine only the product of the $NNv$ coupling 
constants and the form factor.  Consequently one cannot extract a 
unique value for $g_{NNv}$ directly from the analysis; further
constraints---such as the energy dependence of the angular 
distribution---are needed. As discussed in \cite{Nak1}, the energy 
dependence will impose some constraint on the form factor. Also, in the energy
region far from the threshold, the tensor-to-vector coupling ratio $\kappa_v = 
f_{NNv}/g_{NNv}$ influences the energy dependence of the total cross section. 
Another constraint may be imposed by spin polarization observables. In the case of 
$pp$ bremsstrahlung producing hard photons, it is known that the 
reaction is dominated by the magnetization current, to which the tensor 
coupling of the $NN\gamma$ vertex contributes. Therefore we would expect 
that the spin observables in $pp \rightarrow ppv$ reactions become more
sensitive to the tensor coupling $f_{NNv}$ for energetic vector mesons.

%-------------------
 \section{Application }
%---------------------

As mentioned in the introduction, the determination of the $NN\phi$ coupling 
strength is of special interest in the study of vector-meson production since 
its magnitude is usually associated with the amount of hidden strangeness 
in the nucleon. In this section we utilize our model to obtain some information 
on this coupling strength. However, as pointed out in the previous section, the 
angular distribution alone is not sufficient for determining $g_{NN\phi}$ uniquely.
We therefore choose to perform a combined analysis of $\phi$- 
and $\omega$-meson production; i.e., to use the available data from both reactions
to extract $g_{NN\phi}$. A combined analysis of $\phi$ and $\omega$ production also 
allows us to address the important issue concerning the violation of the OZI rule. 

Before doing so some considerations about the existing experimental data 
are in order. So far only very few precision data of vector-meson production in 
$NN$ collisions are available. First there are total cross sections for the reaction 
$pp\rightarrow pp\omega$ in the energy range $T_{lab} \cong 1.89$ to 1.98 GeV 
from Saclay \cite{Saclay}. In addition, there are angular distributions of the emitted 
meson for both $\omega$- and $\phi$-meson production (although with no absolute 
normalization) and the ratio of the total cross sections $\sigma_\phi / \sigma_\omega 
\equiv \sigma(pp \rightarrow pp\phi) / \sigma(pp\rightarrow pp\omega)$ at $T_{lab} = 
2.85$ GeV from the DISTO collaboration \cite{Disto}. All the other presently available 
data \cite{Flam} are from the 1970's and not very accurate. Furthermore these data lie 
in an energy range far above the vector-meson production thresholds. At the 
corresponding excess energies the $NN$ interaction in the final state is already 
dominated by inelastic processes. Such processes are not accounted for in the $NN$ model 
that we employ, which is valid only for energies below the pion-production threshold 
(i.e., excess energies below $\approx 140$ MeV), and therefore we do not consider those 
high-energy data in our analysis. Since the excess energy in the $\omega$-meson 
production channel of the DISTO measurement ($Q = 319$ MeV) is already beyond the energy 
range where we trust our model, we do not use the measured production ratio directly. 
Instead we interpolate the total cross section for $\omega$-meson production from the 
existing data \cite{Saclay,Flam} and use this value, and the measured ratio, to fix the 
absolute normalization of the measured $\phi$-meson angular distribution \cite{Disto}. 
Our estimate yields $\sigma_\omega \sim 0.7 \times 10^2 ~\mu$b at $T_{lab}=2.85$ GeV. 
This value of $\sigma_\omega$, combined with the measured ratio of 
$\sigma_\phi / \sigma_\omega = (3.7 \pm 0.5) \times 10^{-3}$ \cite{Disto}, leads to 
$\sigma_\phi \sim 0.26~ \mu$b. Inevitably, such an interpolation is subject to 
uncertainties. We have, therefore, carried out the same analysis as reported below, but
starting from a $\phi$-production cross section that is smaller/larger by about 40\%. 
We arrived at basically the same conclusions and therefore refrain from showing the 
corresponding results here. 

Since we have only one $\phi$-meson angular distribution and five $\omega$-meson total
cross section data available in the range of applicability of our model, we are forced to 
impose some constraints on the model in order to reduce the number of free parameters. To 
this end we make the following assumptions:

\begin{itemize}
\item[1)] 
The same cutoff parameter $\Lambda_M\equiv \Lambda_{M\omega} = \Lambda_{M\phi}$ 
(cf. Eq.(\ref{formfactorM})) is used for the meson-exchange currents for $\omega$- and 
$\phi$-meson production. This is a reasonable choice since the off-shell particles at 
the $v\rho\pi$ vertex are the same in both production processes. Likewise, the cutoff 
parameter $\Lambda_{Nv}$ (cf. Eq.(\ref{formfactorN})) in the nucleonic current is assumed 
to be the same for $\omega$- and $\phi$-meson production, i.e., $\Lambda_N\equiv 
\Lambda_{N\omega}=\Lambda_{N\phi}$. 

\item[2)] 
The $NN\omega$ vector coupling constant is given by the value obtained from
SU(3) flavor symmetry and imposing the OZI rule, i.e., $g_{NN\omega} = 3 
g_{NN\rho} \cos(\alpha_V)$, where $\alpha_V \equiv \theta_V - \theta_{V(ideal)}$ 
is the deviation from the ideal $\omega - \phi$ mixing angle. With $\alpha_V \cong 
3.7^o$ \cite{PDG} and the $NN\rho$ coupling constant of $g_{NN\rho} = 2.3 - 3.36$ 
\cite{rhocoup}, this yields a value of $g_{NN\omega} \cong (9 \pm 2)$, which is 
close to the value of $g_{NN\omega}\cong 11$ obtained 
in a recent $NN$ scattering analysis
\cite{Janssen}. The uncertainty here comes from the uncertainties involved in 
$g_{NN\rho}$ and $\alpha_V$. 

\item[3)] 
The tensor-to-vector coupling ratio, $\kappa_v = f_{NNv}/g_{NNv}$, is the same
for the $\omega$- and $\phi$-mesons: $\kappa \equiv\kappa_\omega = \kappa_\phi$. 
This relation is also suggested by SU(3) symmetry. Furthermore we choose the parameter 
$\kappa$ to be fixed beforehand. We consider the values $\kappa = \pm 0.5$, which 
covers a rather ample range. Unlike the case of the $\rho$-meson, we do not expect 
$f_{NNv} = \kappa_v g_{NNv}$ to be large, especially for the $\omega$-meson. This is 
supported by an estimate \cite{Dover}, in which SU(3) is applied to the sum
$f_{NNv} + g_{NNv}$ rather than to $f_{NNv}$ alone, which is motivated by the
success of SU(6) in predicting the magnetic moments of the baryons. Other 
investigations \cite{Meissner,MHE87} also support small values of $\kappa_v$.
\end{itemize}

These assumptions reduce the number of parameters of the model to be fixed to 
a total of five: the cutoff parameters $\Lambda_M$ and $\Lambda_N$, and the coupling 
constants $g_{\phi\rho\pi}$, $g_{\omega\rho\pi}$, and $g_{NN\phi}$. As mentioned 
in the previous section, the first two coupling constants can be extracted from 
the measured branching ratios. Specifically, the coupling constant of
$g_{\phi\rho\pi}=-1.64$ is determined directly from the measured decay width of $\phi
\rightarrow \rho + \pi$ \cite{PDG}. The coupling constant $g_{\omega\rho\pi}$, however,
cannot be determined directly (as in the case of $g_{\phi\rho\pi}$) since $\omega 
\rightarrow \rho + \pi$ is energetically forbidden. We therefore extract it 
indirectly from the radiative decay width of $\omega \rightarrow \pi + \gamma$, assuming 
vector meson dominance; we obtain $g_{\omega\rho\pi}=10$. The signs of these 
couplings are inferred from SU(3) symmetry considerations. We note that these coupling 
constants are extracted at different kinematics: $g_{\phi\rho\pi}$ is determined at 
$q^2_\rho = m^2_\rho$ and $q^2_\pi = m^2_\pi$, whereas $g_{\omega\rho\pi}$ is extracted 
at $q^2_\rho = 0$ and $q^2_\pi = m^2_\pi$. The corresponding form factor (cf. 
Eq.(\ref{formfactorM})) should, therefore, be normalized accordingly, i.e., $x=1$ and 
$x=0$ for the $\phi\rho\pi$ and $\omega\rho\pi$ vertex form factor, respectively. With 
the coupling constants at the three-meson-point vertices fixed, we are then left with 
three free parameters which may be adjusted to reproduce the $\phi$- and $\omega$-meson 
production data.

\begin{figure}[h]
\vspace{10cm}
\includegraphics{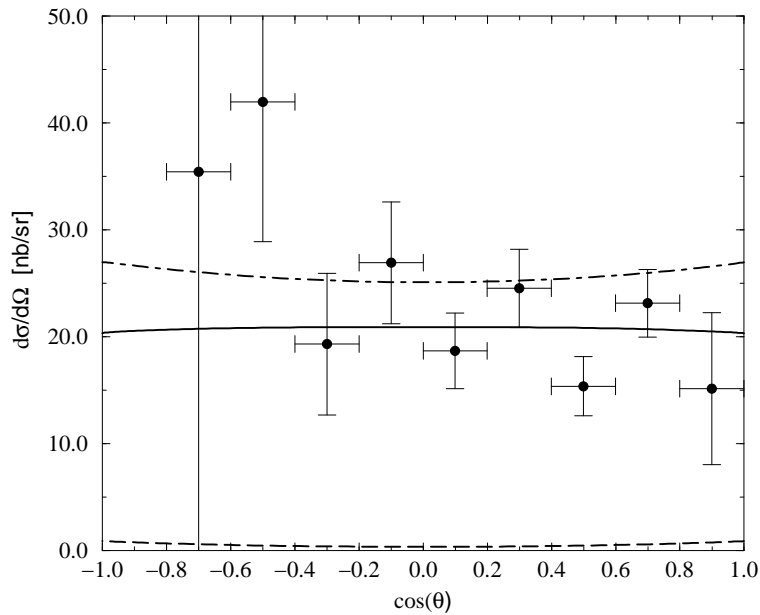}
\caption{Angular distribution for the reaction $pp\rightarrow pp\phi$ 
at an incident energy of $T_{lab}=2.85$ GeV. The dashed-dotted curve corresponds 
to the mesonic current contribution, the dashed curve to the nucleonic 
current contribution. The solid curve is the total contribution. The experimental 
data are from Ref.~\protect\cite{Disto} and have been normalized according to the 
procedure explained in the text.}
\label{fig2}
\end{figure}

\begin{figure}[h]
\vspace{10cm}
\includegraphics{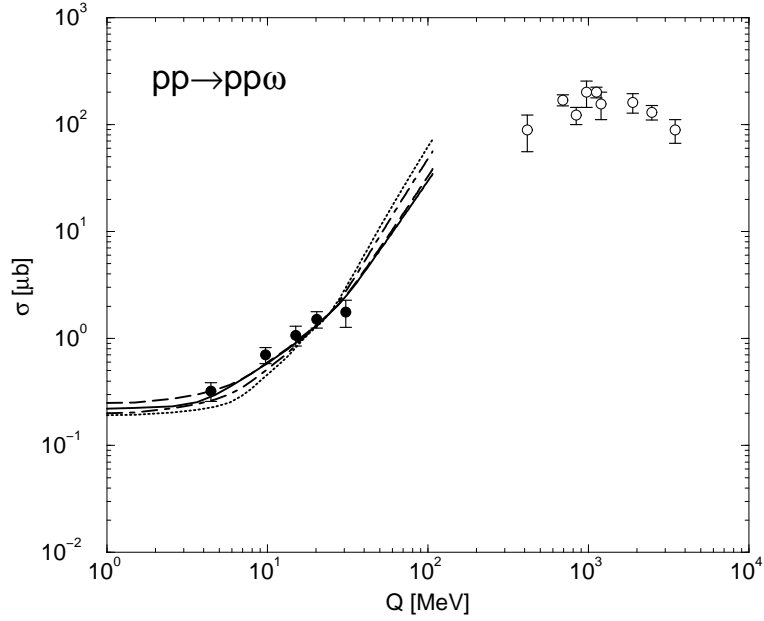}
\caption{Total cross section for the reaction $pp \rightarrow pp\omega$ as a 
function of the excess energy $Q=\sqrt{s}-\sqrt{s_o}$. The dotted, dash-dotted, 
dashed, and solid lines correspond to the model calculation based on the cutoff 
mass of $\Lambda_N$ = 1545, 1312, 1411, and 1170 MeV, respectively, as given in 
Table~\protect\ref{tab2}. Effects of the $\omega$-meson mass distribution are 
taken into account. The experimental data are from Ref.~\protect\cite{Saclay}
(filled circle) and from Ref.~\protect\cite{Flam} (open circle).}
\label{fig3}
\end{figure}

We are now prepared to apply the model to the reactions $pp\rightarrow pp\omega$ and 
$pp\rightarrow pp\phi$. Before doing any calculation, however, we note that the 
nucleonic current contribution to the $\phi$-meson production should be rather small, 
as the measured angular distribution shown in Fig.~\ref{fig2} is more or less 
isotropic. Recall that the angular distribution alone is sufficient to establish the 
magnitude of both the nucleonic and mesonic currents uniquely, and that 
the mesonic current yields a flat angular distribution, whereas the nucleonic current 
gives an approximately $ \cos^2(\theta)$ dependence~\cite{Nak1}. For the determination of the free 
parameters $\Lambda_M$, $\Lambda_N$ and $g_{NN\phi}$ we proceed in the following way: 
we assume that the angular distribution of the $\phi$-meson resembles the solid curve 
in Fig.~\ref{fig2}. We then determine the required contributions from both the nucleonic
and mesonic currents. The mesonic current involves only one free parameter: namely the 
cutoff mass $\Lambda_M (\equiv\Lambda_{M\phi}=\Lambda_{M\omega})$ of the $\phi\rho\pi$ vertex 
form factor (cf. Eq.(\ref{formfactorM})). That is fixed by the requirement of reproducing 
the angular distribution.  We obtain $\Lambda_M = 1450$ MeV.  Turning now to $\omega$ 
production, we require that our model describe the total cross section data from SATURNE 
\cite{Saclay}. Since we assumed the $NN\omega$ vector coupling to be $g_{NN\omega} = 9$ (and
$\kappa\equiv\kappa_\omega=\kappa_\phi$ is fixed to the two values mentioned above), one can 
determine the cutoff parameter $\Lambda_N (\equiv\Lambda_{N\omega}=\Lambda_{N\phi})$ in 
Eq.(\ref{formfactorN})---the only remaining free parameter---from the $\omega$-production 
cross section. Owing to the destructive interference between the nucleonic and mesonic 
currents, we find, in general, two possible values of $\Lambda_N$ for given $g_{NN\omega}$ 
and $\kappa$. The resulting values are listed in Table~\ref{tab2} and the corresponding cross 
sections are shown in Fig.~\ref{fig3}. As pointed out in \cite{Nak1}, in contrast to the 
angular distribution, the total cross section is unable to constrain uniquely the absolute 
contribution of the nucleonic and mesonic currents. As can be seen from Fig.~\ref{fig3}, 
the energy dependence of the total cross section may be useful to further constrain the 
ratio $\kappa$ and/or the form factor at the production vertex. Unfortunately, no 
data is available at energies that are well above the threshold, yet still low enough for 
the present model to be applicable. It should be mentioned that effects of the finite 
width of the $\omega$-meson---which influence considerably the energy dependence of the 
total cross section close to the threshold energy \cite{Wilkin}---haven been taken into 
account in the results shown in Fig.~\ref{fig3}. This is done by folding the calculated 
cross section with the Breit-Wigner mass distribution of the $\omega$-meson. Once the 
cutoff parameter $\Lambda_N$ is fixed, one can return to the angular distribution of the 
$\phi$-meson production in Fig.~\ref{fig2} and adjust the $NN\phi$ coupling constant
$g_{NN\phi}$ to reproduce the amount of the nucleonic current contribution previously 
determined. Since we have two values of $\Lambda_N$ for each chosen value of $\kappa$, 
we get four values of $g_{NN\phi}$ corresponding to the four possible combinations. The
values for the $NN\phi$ coupling constant thus extracted are compiled at the bottom of 
Table~\ref{tab2}. They range from $g_{NN\phi}=-0.19$ to $g_{NN\phi}=-0.90$. As already 
acknowledged above, the lack of a more complete and accurate set of data prevents
us from achieving a more accurate determination of $g_{NN\phi}$.

The values of $g_{NN\phi}$ thus obtained may be compared with those resulting 
from SU(3) flavor symmetry considerations and imposition of the OZI rule,
\begin{equation} 
g_{NN\phi} = - 3 g_{NN\rho} \sin(\alpha_V) \cong  -(0.60 \pm 0.15) \ ,
\label{OZI}
\end{equation}
where the factor $\sin(\alpha_V)$ is due to the deviation from the ideal $\omega - \phi$ 
mixing. The numerical value is obtained using the values  $g_{NN\rho} = 2.63 - 3.36$ 
\cite{rhocoup} and $\alpha_V \cong 3.7^o$ \cite{PDG}. The error bar quoted is due to the 
uncertainty involved in $g_{NN\rho}$ and $\alpha_V$. Comparing the value (Eq.(\ref{OZI})) 
with those obtained in the present work we conclude that the $\phi$-production data can be 
described with $NN\phi$ coupling constants that are compatible with the OZI value.

It should be mentioned that although our analysis of the existing DISTO and SATURNE data 
yields $NN\phi$ coupling constants which are compatible with the OZI value, it is necessary to 
introduce a violation of the OZI rule at the $v\rho\pi$ vertices ($v=\phi , \omega$) in the
meson-exchange current. This can easily be verified by calculating, e.g., the $\phi\rho\pi$
coupling constants at $q_\rho^2=0$ using the form factor given by Eq.(\ref{formfactorM})
with the cutoff parameter $\Lambda_M$ extracted from the $\phi$-meson angular distribution 
data. We obtain $g_{\phi\rho\pi}(q_\rho^2=0) = g_{\phi\rho\pi}(q_\rho^2=m_\rho^2)\times 
F_{v\rho\pi}(q_\rho^2=0,q_\pi^2)=-1.18$. This value is almost a factor of 2 larger than the 
corresponding OZI value of $g_{\phi\rho\pi}(q_\rho^2=0) = - g_{\omega\rho\pi}(q_\rho^2=0)
\tan(\alpha_V) \cong -0.65$ given in Table~\ref{tab1}. We found it impossible to 
describe the data without introducing this OZI violation (independently of the considerations 
about the $NN\phi$ coupling constant) for the following reasons: A sufficiently large 
contribution from the mesonic current in the $\phi$-meson production, as demanded by the 
measured angular distribution, can only be obtained with the $\phi\rho\pi$ coupling constant 
extracted from the measured decay width of $\phi \rightarrow \rho + \pi$ \cite{PDG}, i.e., 
$g_{\phi\rho\pi} = -1.64$. For the coupling constant extracted from radiative meson decay, an 
unrealistically large cut-off mass $\Lambda_M$ in excess of $3 GeV$ would be required. On the 
other hand, with $g_{\phi\rho\pi} = -1.64$ and the corresponding OZI value of $g_{\omega\rho\pi} 
= -g_{\phi\rho\pi}/ \tan(\alpha_V) = 25$ for the $\omega\rho\pi$ coupling constant, we have 
little chance of describing the energy dependence of the $\omega$-meson total cross section; it 
simply rises much too strongly with the energy. In this context let us mention that a knowledge 
of the angular distribution of the $\omega$-meson in the energy region where the 
model is applicable is of particular importance. Corresponding data would enable 
us to pin down the contributions for the mesonic current and thereby 
impose much more stringent constraints on the parameters of the model relevant 
to a possible violation of the OZI rule at the $v\rho\pi$ vertices. 
It would be very useful to clarify this point, specifically
because an analysis of the radiative decay widths of the vector mesons within 
the vector-meson dominance model, involving the very same coupling constants,
led to $v\rho\pi$ coupling constants which satisfy the OZI rule to within 
$15-20\%$ \cite{Durso}.
Furthermore it is certainly desirable to test the present model
in describing other independent reactions processes (e.g., $\phi$- and
$\omega$-meson photoproduction). We plan such investigations in the
near future.

Finally, we wish to make a remark on the ``naive'' OZI estimation as expressed 
by Eq.(\ref{Ratio}). This equation is simply a consequence of assuming that the 
$\phi$- and $\omega$-meson production amplitudes differ from each other only by their 
coupling strengths, which are assumed to be related by SU(3) flavor symmetry plus the 
OZI rule. Differences in the kinematics (besides trivial phase-space effects) 
induced, for example, by the mass difference between the $\omega$- and 
$\phi$ mesons, interference effects between the nucleonic and meson-exchange currents, etc., 
are completely ignored. Our model offers the possibility to test explicitly the 
validity of this assumption and thus the reliability of Eq.~(\ref{Ratio}). For this purpose
we carried out a (full) model calculation where we imposed the OZI rule to relate the 
relevant coupling constants; i.e., $g_{\phi\rho\pi} / g_{\omega\rho\pi} = - \tan(\alpha_V)$ 
and $g_{NN\phi} / g_{NN\omega} = - \tan(\alpha_V)$. In addition, we used the same form factors 
at the $\omega$- and $\phi$-meson production vertices in the mesonic and the nucleonic current,
respectively. It turned out that such a calculation yields results that are qualitatively very 
similar to the simple estimate of Eq.(\ref{Ratio}), indicating that the above-mentioned 
differences in the kinematics, etc., do not induce any significant deviation from 
Eq.(\ref{Ratio}). In fact, the cross section ratio taken at the same excess energy is roughly 
$3 \times 10^{-3}$, or about 30\% smaller than the ''naive`` OZI estimate.

%----------------
\section{Summary}
%----------------

We have investigated the reaction $pp \rightarrow pp\phi$ using a relativistic 
meson-exchange model. We find that the nucleonic and $\phi\rho\pi$ exchange 
currents are the two dominant sources contributing to $\phi$-production in 
this reaction, and that they interfere destructively. 
Since these two reaction mechanisms give rise to distinct angular 
distributions, measurements of this observable can yield valuable 
information on the magnitude of the nucleonic current and, specifically,
on the $NN\phi$ coupling constant. In fact, the flat angular 
distribution exhibited by the data from DISTO collaboration \cite{Disto} 
indicates that the contribution from the nucleonic current must be very small. 
This, in turn, implies that the value of the $NN\phi$ coupling constant must 
also be small. Indeed, our semi-quantitative combined analysis of those
$\phi$-meson production data and the data for $\omega$-meson 
production from SATURNE yields values of $g_{NN\phi}$ which are compatible 
with the OZI rule.

{\bf Acknowledgments}

J.D. acknowledges
the hospitality of the Institute f\"ur Kernphysik, Forschungzentrum
J\"ulich. One of the authors (C.H.) is grateful for financial support by
the COSY FFE--Project No. 41324880.

\vfill \eject

\newpage

\begin{table}
\caption{SU(3) based estimate of the $\phi VP$ coupling constants,
         $g_{\phi VP}$ at the vanishing four-momentum square of the
         vector-meson $V$. The coupling constants are given in units 
         of $1/\sqrt{m_\phi m_V}$, with $m_V$ denoting the mass of the 
         vector-meson $V$ involved. The parameters of the SU(3) 
         effective Lagrangian are taken from Ref.~\protect\cite{Durso} 
         (model B).}
\vskip -0.25cm
\tabskip=1em plus2em minus.5em
\halign to \hsize{\hfil#\hfil&\hfil#\hfil&\hfil#\hfil
                 &\hfil#\hfil&\hfil#\hfil\cr
\noalign{\hrulefill}
 $g_{\phi\rho\pi}$ & $g_{\phi\phi\eta}$ & $g_{\phi\omega\eta}$ & 
$g_{\phi\phi\eta '}$ & $g_{\phi\omega\eta '}$ \cr
\noalign{\vskip -0.20cm}
\noalign{\hrulefill}
\noalign{\vskip -0.5cm}
\noalign{\hrulefill}
 -0.65 &  9.86 &  0.19 & -9.80 & -1.11 \cr
\noalign{\hrulefill} }
\label{tab1}
\end{table}
\begin{table}
\caption{$NN\phi$ coupling constant extracted from our model analysis of
the reaction $pp\rightarrow ppv$ ($v=\phi ,\omega$) as described in Sect. III
for two given values of the ratio $\kappa = f_{NNv}/g_{NNv}$.}
\vskip -0.25cm
\tabskip=1em plus2em minus.5em
\halign to \hsize{\hfil#\hfil&\hfil#\hfil&\hfil#\hfil&
                  \hfil#\hfil&\hfil#\hfil\cr
\noalign{\hrulefill}
$\kappa$      & -0.5    &   -0.5  &  +0.5   &  +0.5   \cr
\noalign{\hrulefill}
$\Lambda_N$   &  1170   &   1411  &  1312   &  1545    \cr
\noalign{\hrulefill}
$g_{NN\phi}$  &  -0.45  &  -0.19  &  -0.90  &  -0.40   \cr
\noalign{\hrulefill} }
\label{tab2}
\end{table}

\newpage

\newpage
\vfill 2cm

\end{document}